\documentclass[preprint,5p,times,twocolumn]{elsarticle}
\pdfoutput=1

\usepackage{amssymb}

\usepackage{algorithm}
\usepackage{algorithmic}
\usepackage{mathtools}
\usepackage{tabularx}
\usepackage{stfloats}
\usepackage{hyperref}
\usepackage{fancyvrb}
\usepackage{xcolor}
\usepackage{soul}
\usepackage{subcaption}
\VerbatimFootnotes

\journal{Journal of Computational Physics}

\begin{document}

\begin{frontmatter}


\author[inst1]{Stefan Lüders\corref{cor1}}
\ead{psi@stefan-lueders.de}
\cortext[cor1]{Corresponding Author}
\author[inst1]{Klaus Dolag}
\affiliation[inst1]{organization={University Observatory Munich},
             addressline={Scheinerstr. 1},
             city={Munich},
             postcode={D-81679},
             state={Bavaria},
             country={Germany}}

\title{PSI: Constructing ad-hoc Simplices to Interpolate High-Dimensional Unstructured Data}




\begin{abstract}
Interpolating unstructured data using barycentric coordinates becomes infeasible at high dimensions due to the prohibitive memory requirements of building a Delaunay triangulation. We present a new algorithm to construct ad-hoc simplices that are empirically guaranteed to contain the target coordinates, based on a nearest neighbor heuristic and an iterative dimensionality reduction through projection. We use these simplices to interpolate the astrophysical cooling function $\Lambda$ and show that this new approach { produces good results with just a fraction of the previously required memory.}
\end{abstract}




\end{frontmatter}


\newpage

\section{Introduction}
\label{sec:introduction}

The so-called \textit{cooling function} $\Lambda$ describes the energy loss of a gas cloud per unit volume and time and is an important consideration in cosmological simulations (e.g.\ \cite{draine:2011}).
Besides density and temperature, $\Lambda$ depends on many additional, local properties, such as the chemical composition of the gas cloud and various spectral energy distributions that shape the radiation background. { As a result, $\Lambda$ is time consuming to compute and has a complicated functional form. This includes large gradients that can span more than three orders of magnitude. At the same time a cosmological simulation may require its calculation on the order of a billion times or more.}
As such, $\Lambda$ is usually interpolated from regular grids of { precalculated values, which typically omit higher dimensions to avoid memory issues and only provide approximative results.}

In \cite{lueders:21} we improved on this with the {Cloudy based Heuristic and Iterative Parameterspace Sampler (CHIPS,}~ \url{https://github.com/Vetinar1/CHIPS}) that eschews regular grids in favor of amorphous sample distributions that take the shape of $\Lambda$ into account. This both reduced the number of required samples and increased the interpolation accuracy. There exist a variety of methods for interpolat{ing} the resulting unstructured data \cite{hoschek:92}. {We achieved the best results using a simple Delaunay tessellation based method} that executes a directed search for the simplex containing the target point and then interpolates it using barycentric coordinates\cite{moebius:1827}. This resulted in our Delaunay Interpolation Program DIP (\url{https://github.com/Vetinar1/DIP}).

{In cosmology, tessellation based interpolation has been previously applied to velocity fields and density distributions (e.g.~\cite{bernardeau:1996}, \cite{weygaert:2001}, \cite{schaap:2007} \cite{Cautun:2011}). The simplest approach is Voronoi Tessellation Field Estimation (VTFE), in which the functional value of a point is used to approximate the function within the entire corresponding cell. This zeroth-order interpolation is equivalent to a nearest neighbor estimation and produces sufficient approximations for slow-changing functions, but is not satisfactory for the highly varied cooling function.

In contrast, Delaunay Tessellation Field Estimation (DTFE) is a first order approach, equivalent to linear interpolation in higher dimensions. Unlike VTFE it does not produce discontinuities at cell edges. DIP is similar to existing DTFE software \cite{Cautun:2011}, but does not consider the location of the points themselves as data, and has been developed specifically to be integrated in cosmological simulation software.}

The main downside of {a Delaunay based approach} is the space complexity of the triangulation which scales strongly with sample count and dimensionality, making it infeasible for high dimensional parameter spaces. { While space efficient Delaunay algorithms such as \Verb@Del_graph@ \cite{boissonnat:2009} can alleviate this issue, they can not completely remedy it, as sample counts may reach millions depending on the desired accuracy.} Since building small-scale triangulations near the target is impractical due to the associated runtime, we developed a novel simplex construction heuristic that can be used instead, allowing for much higher sample counts and dimensions. We call it the Projective Simplex Algorithm and implement it in the Projective Simplex Interpolation (PSI) package which is made available as part of DIP.

\section{Description of the Algorithm}
\label{sec:description}

Let $f$ be some function known at points $P \in \mathbb{R}^D$ and let the target point $t$ be a point within the convex hull of $P$. Our goal is to approximate $f(t)$ without building a triangulation on $P$ or a subset of $P$.

We apply the following three step process:

\begin{enumerate}
    \item Identify $P_k$, the $k$ nearest neighbors of $t$ in $P$.
    \item Construct a simplex $S$ out of the points in $P_k$ such that $t$ is contained within $S$.
    \item Calculate the barycentric coordinates of $t$ relative to $S$ and use them to interpolate.
\end{enumerate}

Finding the $k$ nearest neighbors of a given point is a well researched problem and we chose to use ball trees for this purpose, see \cite{omohundro:1989} for further information.

The barycentric coordinates $\lambda_i$ of a simplex are homogeneous coordinates relative to its $D+1$ vertices. They refer to that point in space which would be the center of mass (``barycenter") of the simplex if each vertex $v_i$ had the mass of the associated barycentric coordinate $\lambda_i$. If their sum is normalized to unity they can be used to linearly interpolate on the $D$-dimensional hyperplane spanned up by the simplex. A good introduction on the subject can be found in \cite{vince:2017}.

The contribution of this paper is a new heuristic to directly select points for the simplex, the Projective Simplex Algorithm. The underlying idea is that the nearest neighbor of the target point is likely to be part of a small, high quality simplex containing $t$.
The dimension along the vector between $t$ and the nearest neighbor is then eliminated through projection and the problem is reduced to an equivalent but lower dimensional one.
Algorithm \ref{alg:psa} shows the steps of the algorithm in detail.

\begin{algorithm}
\caption{The Projective Simplex Algorithm.}
\label{alg:psa}
\begin{algorithmic}[1]
\REQUIRE Set of points $P_k$ in $\mathbb{R}^D$, target point $t$, $S = \emptyset$
\STATE Let $d$ equal the number of dimensions
\STATE $\mathcal{P} := P_k$, $t' := t$
\WHILE{$d > 1$}
    \STATE Let $a$ be the nearest neighbor of $t'$ in $\mathcal{P}$
    \STATE $S \leftarrow S \cup \{p \in P_k \mid a \text{ is projection of } p \}$
    \STATE $\mathcal{P} \leftarrow \mathcal{P} \setminus \{a\}$
    \STATE $\vec{n} := a - t'$, $\mathcal{P}' := \emptyset$
    \FORALL{$\vec{p} \in \mathcal{P}$}
        \STATE Project $\vec{p}$ on $\vec{n}$: $\vec{p}_\perp := \frac{\vec{p} \cdot \vec{n}}{\vec{n} \cdot \vec{n}} \cdot \vec{n}$
        \STATE Shift to origin: $\vec{p}_{\perp,0} := \vec{p}_\perp - t'$
        \IF{$\vec{p}_{\perp,0} \cdot \vec{n} < 0$}
            \STATE Add projected point: $\mathcal{P}' \leftarrow \mathcal{P}' \cup \{\vec{p} - \vec{p}_\perp\}$
        \ENDIF
    \ENDFOR
    \STATE Project target point: $t' \leftarrow t' - \frac{t' \cdot \vec{n}}{\vec{n} \cdot \vec{n}} \cdot \vec{n}$
    \STATE $\mathcal{P} \leftarrow \mathcal{P}'$
    \STATE $d \leftarrow d - 1$
\ENDWHILE
\STATE The remaining projected points are on a line in $\mathbb{R}^D$. Add the nearest neighbor in each direction on the line to $S$, if they exist
\RETURN $S$
\end{algorithmic}
\end{algorithm}

\begin{figure*}[h]
    \centering
    \begin{subfigure}{0.24\textwidth}
        \includegraphics[width=\linewidth]{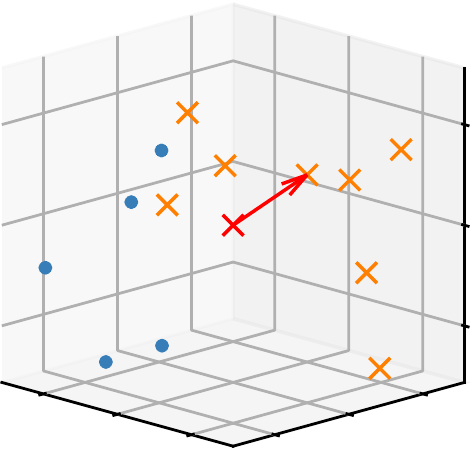}
        \caption{}
    \end{subfigure}
    \begin{subfigure}{0.24\textwidth}
        \includegraphics[width=\linewidth]{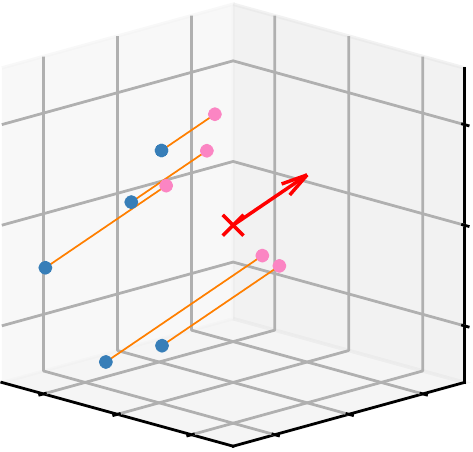}
        \caption{}
    \end{subfigure}
    \begin{subfigure}{0.24\textwidth}
        \includegraphics[width=\linewidth]{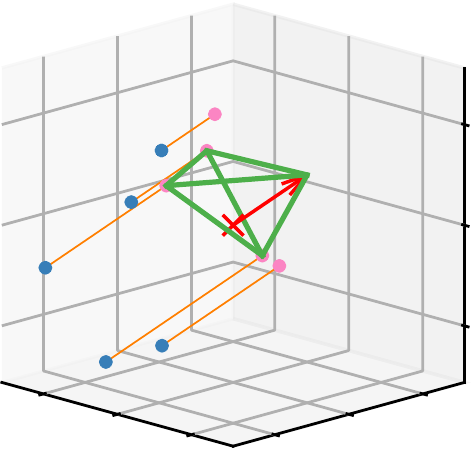}
        \caption{}
    \end{subfigure}
    \begin{subfigure}{0.24\textwidth}
        \includegraphics[width=\linewidth]{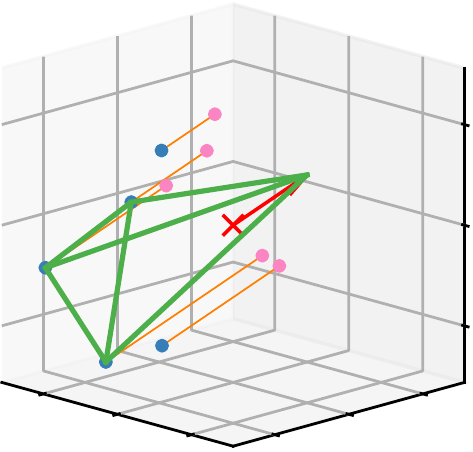}
        \caption{}
    \end{subfigure}
    \caption{This 3-dimensional example illustrates the core idea of the Projective Simplex Algorithm. (a) The red arrow represents a vector pointing from the target point $t$ to the nearest neighbor. Interpreted as a normal vector it defines a plane. Points below this plane remain (blue dots), while points above the plane are removed (orange crosses). (b) The remaining points are projected into the plane (pink points). (c) We can construct a simplex containing $t$ in $\mathbb{R}^3$ by finding a triangle in the projection plane that contains $t$ and adding the nearest neighbor to form a ``pyramid''. The problem of finding this triangle in the 2-dimensional subspace of the projection plane is equivalent to the original problem of finding a simplex. (d) If the simplex found using the projections in (c) contains $t$ then the simplex using the original points represented by these projections must also contain $t$ since all the original points are below the plane.}
    \label{fig:alg}
\end{figure*}

First, working copies $\mathcal{P}$ and $t'$ are made that the algorithm can modify in-place. We find the current nearest neighbor of $t'$ in $\mathcal{P}$ and add it to our simplex $S$. It is removed from further considerations.

Next, the vector $\vec{n}$ pointing from $t'$ to the nearest neighbor is calculated. This vector is the normal vector defining a $(d-1)$-dimensional hyperplane through $t'$ that splits $\mathcal{P}$ into two sets. The points ``above" the plane are in the half space that contains the nearest neighbor and the normal vector. All other points are considered to be below the plane.

If $\vec{n}$ points to the tip of a pyramid-like simplex $\mathcal{S}_d$ with its base inside the plane, then this base face is itself a $(d-1)$-dimensional simplex $\mathcal{S}_{d-1}$ consisting of $d$ points. If $\mathcal{S}_{d-1}$ contains $t'$ then $t'$ is on the surface of $\mathcal{S}_d$, which we consider to be ``inside''. It is now easy to see that if any of the vertices of the base face were moved perpendicularly below the plane, the resulting $\mathcal{S}_d'$ would still contain the target point, since this deformation can never lead to any of the faces crossing $t'$.

This property is the core idea of the algorithm. However, instead of picking points on the plane and then moving their position, the points in $\mathcal{P}$ are projected onto the plane and selected to form a valid simplex face containing $t'$.

First, each $p \in P$ is projected on $\vec{n}$, giving the part of $p$ perpendicular to the plane. It is shifted to the origin in order to compare it to $\vec{n}$. If the scalar product with $\vec{n}$ is positive, $p$ is above the plane and is discarded. If it is negative the projection is completed and the projected point is added to a new set $\mathcal{P}'$. After all points have been processed $\mathcal{P}$ is updated to $\mathcal{P}'$. Finally, $t'$ is also updated through projection.

Now that the dimension along $\vec{n}$ has been eliminated, the goal is to select $d$ points on the $(d-1)$-dimensional projection plane such that they contain $t'$, i.e.\ to find a simplex containing $t'$ using the points in $\mathcal{P}$. This is equivalent to the original problem and thus the proceess can repeat. {Figure \ref{fig:alg} visualizes these steps using a small example distribution.}

At $d = 1$ all points are on a line in $\mathbb{R}^D$ and the procedure can not be applied anymore. However, the simplex can be completed easily by finding the nearest neighbor in either direction on this line, e.g.\ by taking the scalar product of all difference vectors $p_i - t'$ and a vector parallel to the line, then picking the $p_i$ with the absolute smallest positive and negative result.

Sometimes one or both of these neighbors may not exist. In general, if there are less than $d+1$ points left after the projection step the algorithm should abort or restart with a larger $k$. There are several reasons why this might occur:

\begin{itemize}
    \item If $t$ is near the edges of $P$ there may not be enough points in the vicinity for the algorithm to work at all.
    \item Heterogeneous data may lead to the neighbors not being distributed evenly enough around $t$.
    \item $k$ is too small.
\end{itemize}

The first case can be avoided by padding the edges of the point cloud. The optimal amount of padding depends on the sampling density.

If heterogeneities are an issue one might select the nearest neighbors using a radius criterion instead of choosing the $k$ nearest.

In the latter two cases increasing $k$ and re-running the algorithm usually leads to a valid solution\footnote{We also tried ``rewinding" the algorithm and choosing the second nearest neighbor for the simplex, in order to get a better split. This did not significantly improve results.}. The size of $k$ carries a tradeoff between simplex quality and runtime, where smaller $k$ leads to better simplices but larger runtimes due to re-runs. We recommend doubling $k$ after each failed run and choosing the initial value such that {on average} the algorithm runs 1.2 - 1.5 times.

{ In case neither of these approaches succeed, an alternative way of choosing the first vertex can eliminate most of the remaining failures, at the cost of relaxing the nearest neighbor condition. Instead of choosing the nearest neighbor, choose that point for which the most points remain in the point set after the first iteration. This can be achieved by determining the mean difference vector between the target point and its neighbors, which describes the overall directional bias of the neighbor distribution, and using it to select the point that lies farthest in the opposite direction.

This approach incurs additional runtime cost (but no additional complexity) over the regular algorithm, and leads to slightly poorer numerical results. It is, however, more reliable. We thus recommend using this option as a backup. Choosing additional vertices this way does not seem to provide any further advantages.}

Thus, the algorithm is not guaranteed to find a solution. However, according to our testing the solutions it does find are guaranteed to be valid (i.e.\ contain $t$), and the { simplex construction} succeeds in the overwhelming majority of cases (see \ref{sec:performance}).

Overall, the algorithm has a time complexity of $O(kD^2)$. Projecting and filtering the points takes $kD$ operations, and finding the nearest neighbor using brute force also takes $kD$ operations\footnote{Building a $k$d-tree here did not improve performance since $k$ is usually small, see table \ref{tab:data}.}. Both of these are executed $D$ times. In practice, however, significant time can be saved in the first iteration since the nearest neighbor is already known from finding $P_k$.
Also, $k$ merely represents an upper limit since the number of points in the working copy reduces each iteration.
To find the barycentric coordinates of $t$ after the simplex has been constructed, a linear equation system needs to be solved. This adds an additonal term $O(D^3)$. The time required for finding $P_k$ with the ball tree is negligible. The space complexity for saving $n$ $D$-dimensional points is $O(Dn)$.

By comparison a directed search through a full Delaunay triangulation has a time complexity of $O(FD^3)$, where $F$ is the dimension dependent number of simplices evaluated before the simplex containing $t$ is found. In general, $F << k$ for all $D$ if the initial simplex is chosen well.
The space complexity of a Delaunay triangulation is at least $O(Dn^{\lceil D / 2 \rceil})$.

\section{Performance of the Algorithm}
\label{sec:performance}

We implemented the algorithms described in the previous section in the Projective Simplex Interpolation program (PSI). We compare the performance of PSI with our previous Delaunay Interpolation Program (DIP). To this end we run both algorithms on different data sets of varying dimensionalities\footnote{All data sets available at \url{https://www.usm.uni-muenchen.de/~dolag/Data/PSI}}.

\begin{figure}[h]
    \centering
    \includegraphics[width=0.45\textwidth]{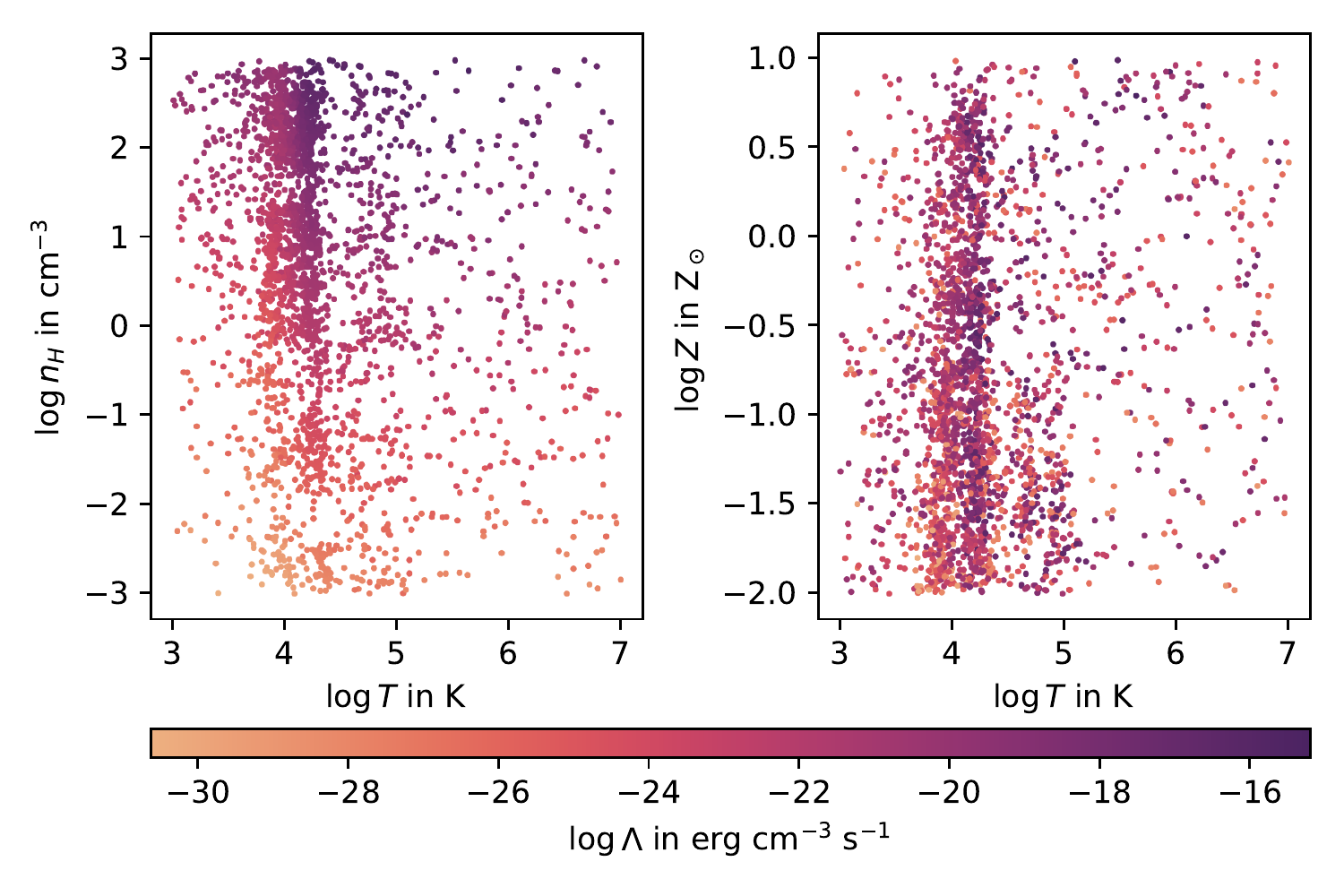}
    \caption{ Two 2D projections of the 3D data set NU\_3D. As a result of the adaptive sampling algorithm CHIPS the hydrogen ionization feature is visible in the point distribution as an overdensity at $T \approx 10^4$K. Interpolating accurately is particularly difficult here. The effects are more pronounced at high densities ($n_H \gtrsim 1\mathrm{cm}^{-3}$) and less pronounced at high metallicities ($Z \gtrsim 0.5Z_\odot$). Note that the values of $\Lambda$ span 14 orders of magnitude.}
    \label{fig:exampledist}
\end{figure}

The non-uniform data sets (NU) were generated using CHIPS\cite{lueders:21} and provide real-world scenarios. They are characterized by an overdense slice of samples through the parameter space corresponding to the ionization temperature of hydrogen {(cf.~fig.~\ref{fig:exampledist})}. We intend to use PSI with {larger versions of} data sets like these. The uniform data sets (U) were generated using a uniform random number generator and are supposed to provide PSI with ideal conditions. Both cover realistic parameter spaces (comparable to \cite{lueders:21}) and contain accurate values for the cooling function $\log \Lambda$, calculated using version 17.02 of \textsc{Cloudy}\cite{cloudy:17}. Typical values for $\log \Lambda$ range from -40 to {-15}. We tested against evaluation sets containing 1000 known values at randomly distributed points.

The metrics we use to evaluate PSI and DIP are memory use, runtime, simplex quality, and interpolation error. There are multiple ways to judge the quality of a simplex (e.g.\ \cite{partha:93}). Here, we use the ratio of the inradius $\rho$ to the longest edge $h_\mathrm{max}$ \cite{george:1998}:
\begin{equation}
\label{eq:quality}
    Q = \alpha \rho / h_\mathrm{max}
\end{equation}
with the dimension dependent normalization factor $\alpha = \sqrt{(2d(d+1))}$. The values of this measure range from 0 for degenerate simplices to 1 for the regular $n$-simplex.

\begin{table*}[b]
\centering
\begin{tabularx}{\textwidth}{lXXXXlllllllX}
Data & $N$   & $S_\mathrm{PSI}$ & $S_\mathrm{DIP}$ & $t_\mathrm{PSI}$ & $t_\mathrm{DIP}$ & $Q_\mathrm{PSI}$ & $Q_\mathrm{DIP}$ & $|\Delta_\mathrm{PSI}|$ & $|\Delta_\mathrm{DIP}|$ & $k$ \\
\hline
NU\_2D   & 489   & 0.02        & 0.04        & 74    & {10}  & 0.53$\pm$0.05  & 0.66$\pm$0.03  & 0.066$\pm$0.008   & 0.061$\pm$0.008   & 15     \\
NU\_3D   & 2692  & 0.11        & 0.78        & 787   & {13}  & 0.37$\pm$0.04  & 0.49$\pm$0.03  & 0.11$\pm$0.02     & 0.098$\pm$0.016   & 50    \\
NU\_4D   & 16950 & 0.83        & 29.8        & 3568  & {55}  & 0.35$\pm$0.03  & 0.45$\pm$0.03  & 0.32$\pm$0.15     & 0.31$\pm$0.16     & 100   \\
NU\_5D   & 26047 & 1.40        & 312         & 12957 & {172} & {0.34$\pm$0.03}  & 0.40$\pm$0.02  & {0.35$\pm$0.15}     & 0.35$\pm$0.15     & {200}   \\
{NU\_6D} & 37519 & 2.1 & 3500  & 12157    & 1296 & 0.31$\pm$0.02  & 0.42$\pm$0.01  & 0.44$\pm$0.18     & 0.42$\pm$0.2     & 250   \\
{NU\_7D} & 67970 & 4.2 & 56000 & 38680    & DNF & 0.31$\pm$0.02  & DNF  & 0.53$\pm$0.23     & DNF     & 350   \\
\hline
U\_2D    & 500   & 0.03        & 0.05        & 68       & {10}        & 0.59$\pm$0.05  & 0.68$\pm$0.03  & 0.066$\pm$0.013   & 0.068$\pm$0.015   & 10   \\
U\_3D    & 2500  & 0.10        & 0.76        & 351      & {15}       & 0.49$\pm$0.04  & 0.57$\pm$0.02  & 0.095$\pm$0.022   & 0.089$\pm$0.017   & 20   \\
U\_4D    & 15000 & 0.69        & 27.7        & 1853     & {65}       & 0.42$\pm$0.03  & 0.51$\pm$0.02  & 0.14$\pm$0.04     & 0.13$\pm$0.04     & 40   \\
U\_5D    & 25000 & 1.10        & 314         & 5794     & {197}      & 0.38$\pm$0.02  & 0.47$\pm$0.02  & 0.19$\pm$0.07     & 0.17$\pm$0.05     & 80   \\
U\_6D    & 40000 & 2.90        & 3920        & 12563    & {623}      & 0.35$\pm$0.02  & 0.43$\pm$0.01  & 0.24$\pm$0.10     & 0.22$\pm$0.08     & 160  \\
{U\_7D} & 80000 & 4.6 & 67000  & 38477    & DNF      & 0.32$\pm$0.02  & DNF  & 0.26$\pm$0.1     & DNF     & 250   \\
\end{tabularx}
\caption{Results for DIP and PSI on several uniform and nonuniform data sets. Shows the number of points $N$, the choice of $k$ for PSI, the storage requirements $S$ for both algorithms in megabyte, the runtime $t$ for both algorithms in milliseconds, the resulting simplex qualities according to eq.\ \ref{eq:quality}, and the mean absolute errors of the interpolation $\Delta$. {Data sets NU\_6D and NU\_7D required modifying our previously Delaunay based sampling code to use the projective simplex algorithm.}}
\label{tab:data}
\end{table*}

{Using our data sets,} we found that choosing $k$ such that the algorithm runs between 1.2 and 1.5 times { on average} produces a good tradeoff between runtime and simplex quality. Better quality simplices correspond to higher interpolation quality, but for our data this correlation was weaker than we expected. We set PSI to double $k$ if it failed to build a valid simplex, with a maximum of four doublings. If a valid simplex was found or the maximum was reached, $k$ was reset. {With this configuration all but 20 evaluations found valid simplices (5 in NU\_5D, 2 in NU\_6D, 9 in NU\_7D, 4 in U\_7D).}
Across all runs, PSI did not construct any simplices that did not contain their target point. The results of our tests are shown in Table \ref{tab:data}.


The memory requirements of both algorithms are comparable at low dimensions. However, the space complexity of the Delaunay triangulation becomes an issue past five dimensions. The triangulation alone takes up several gigabytes of data. Since PSI only needs to save the points themselves it is much more efficient here, and can easily scale to both high dimensions and point counts in comparison to DIP.

{Across all} dimensions DIP is consistently faster than PSI by an order of magnitude. It not only loads the triangulation itself into memory, but also caches a lot of intermediate information about the simplices to speed up calculations\footnote{Such as centroids, face-midpoints, transformation matrices, and normal vectors.}. Such caching is not possible with PSI. {This is most pronounced between 3D and 5D. We suspect that at higher dimensions the amount of points removed in each iteration of the projective simplex algorithm counteracts the increasing complexity. However, we have no explanation for the fact that PSI took more time to complete for NU\_5D than NU\_6D. } The runtimes in table \ref{tab:data} do not include the setup time of DIP, which depends on the number of dimensions rather than the number of interpolations, and could be two hours long { or more at 6D and above}. The setup time of PSI is negligible in comparison.

\begin{figure}[h]
    \centering
    \includegraphics[width=0.5\textwidth]{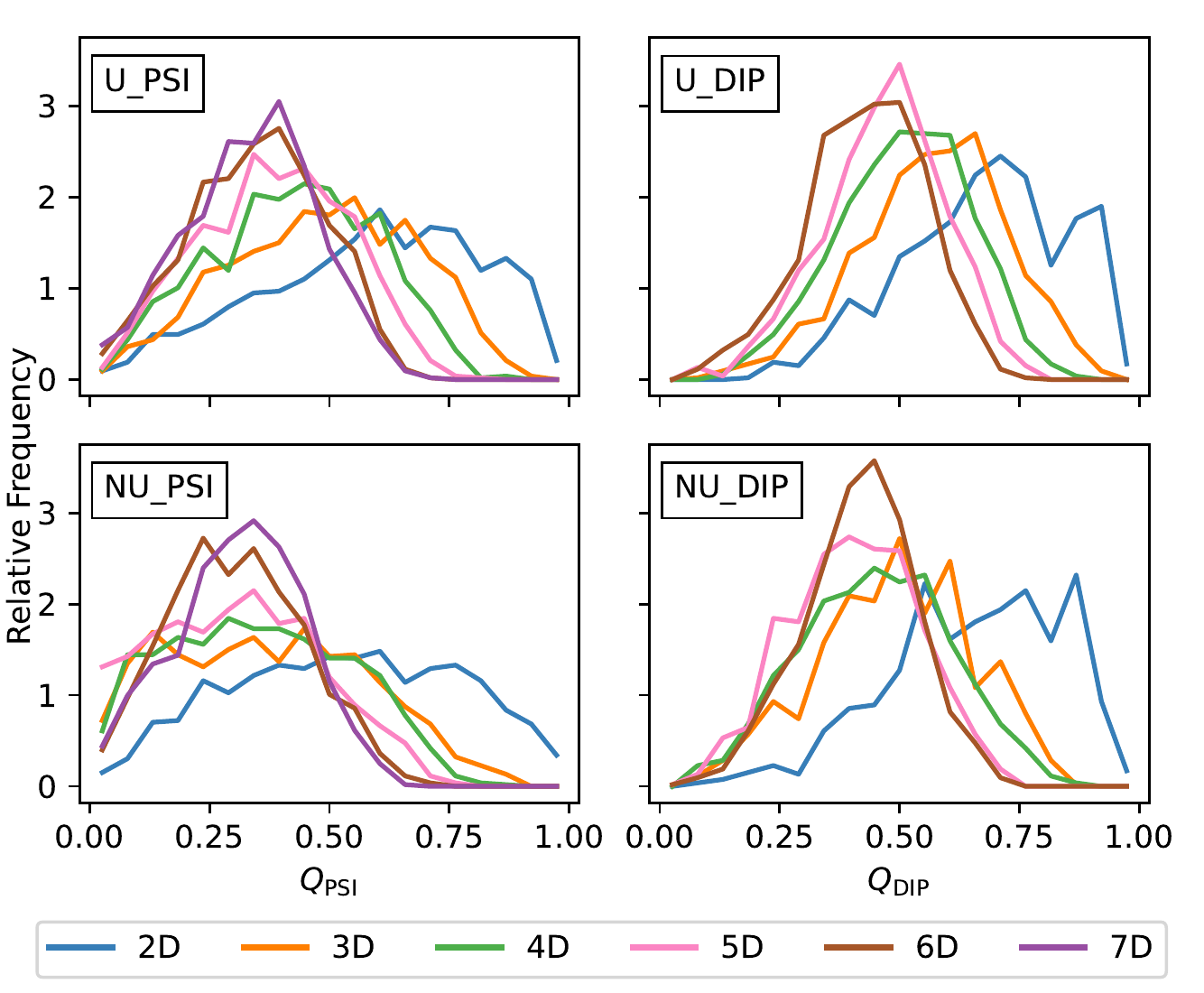}
    \caption{Simplex quality distributions. The upper row shows the uniform data sets while the lower row shows the non-uniform data sets. The left column shows PSI, while the right column shows DIP. Both the Delaunay and PSA simplices suffer in quality as dimension increases.
    }
    \label{fig:qdist}
\end{figure}

The simplex quality is a more abstract measure and is closely related to the interpolation quality. A low quality simplex is long and elongated; using such a simplex to interpolate might produce worse results than a more compact simplex, since the vertices are further away from the target point.
A Delaunay triangulation guarantees an optimal triangulation by maximizing the sum of the interior angles of the simplices, avoiding long thin simplices as much as possible. {The Projective Simplex Algorithm} makes no such guarantee. Instead, distance information is lost through each projection, which can favor elongated simplices particularly for large $k$. As such, the quality of the simplices used in PSI is worse on average and less consistent than the ones used in DIP (cf.\ Fig.\ \ref{fig:qdist}). In the 5D, non-uniform case a worrying number of simplices have a quality of zero. { However, as the dimensionality increases to 7D the mean quality appears to converge to 0.3. DIP shows a similar effect at 0.4. This indicates that PSI could produce reasonable simplices even at much higher dimensions.}

\begin{figure}[t]
    \centering
    \includegraphics[width=0.4\textwidth]{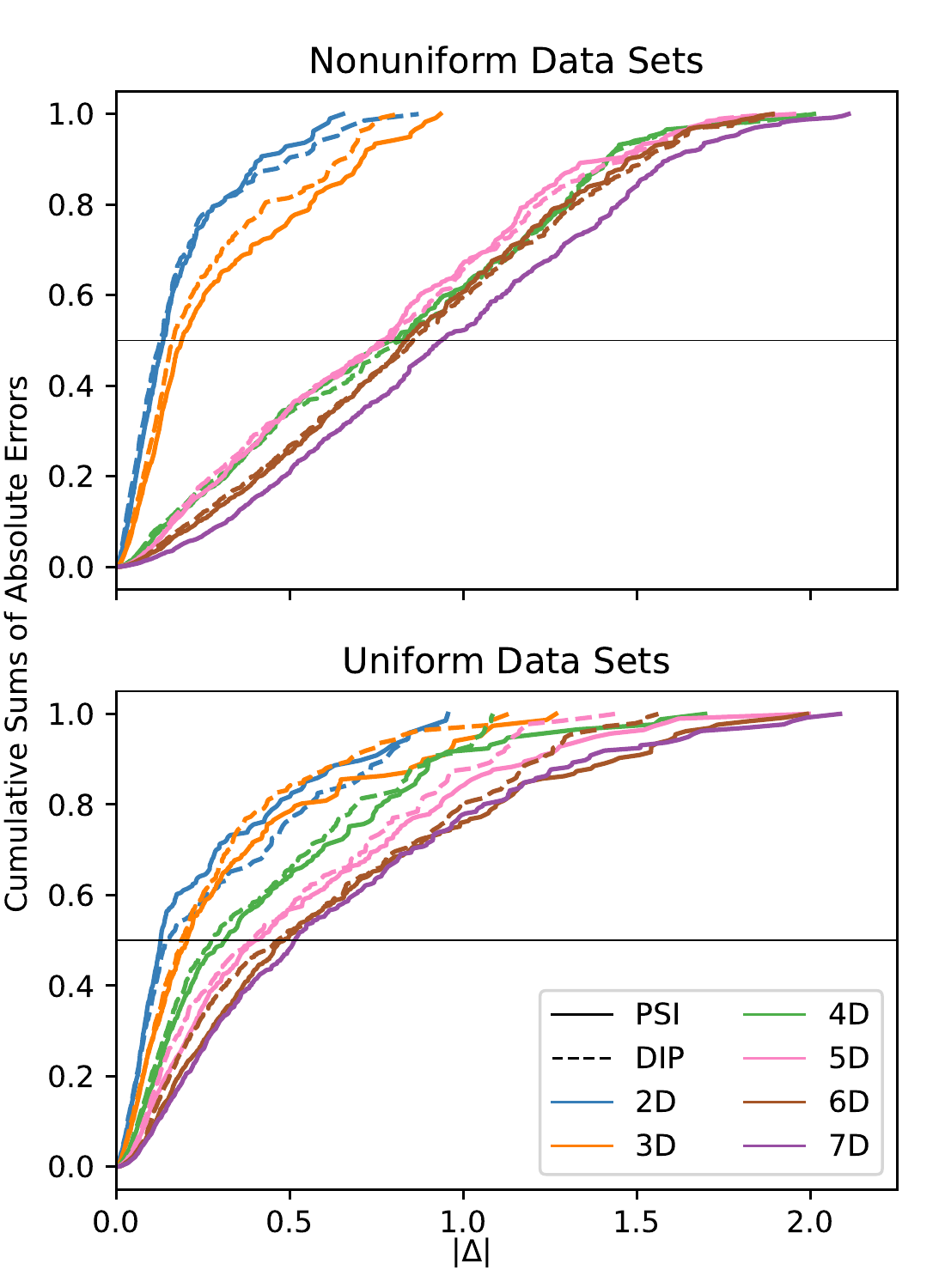}
    \caption{Cumulative sums of the absolute interpolation errors of DIP and PSI.
    }
    \label{fig:cumsums}
\end{figure}

Despite { the lower quality simplices}, the absolute interpolation error $\Delta$ of PSI is similar to the one of DIP. Fig.\ \ref{fig:cumsums} shows the cumulative sums of the error distributions of both DIP and PSI for the different data sets. For the non-uniform data sets PSI performs as well as DIP, except in the 3D case where it performs slightly worse. For the uniform data sets it performs almost as well as DIP in all cases except U\_2D, where it performs even better.
{In practical applications the accuracy can be improved further by increasing sample counts. Due to its lower space complexity PSI can potentially achieve higher accuracy than DIP, which can not support as many samples.}

Overall, PSI performs better than we expected against its predecessor, and we intend to integrate it into our version of the cosmological simulation software \textsc{OpenGadget3}\cite{Springel:2005} soon.

\section{Conclusion}
\label{sec:conclusion}

We developed a new algorithm that constructs a simplex around a target point using an iterative dimensionality reduction through projection, bypassing the need to build a full triangulation. We successfully implemented this algorithm in the Projective Simplex Interpolation program (PSI) and used it to interpolate the cooling function $\Lambda$ needed in galaxy evolution simulations. PSI completely resolved the excessive memory requirements of the previous implementation { with acceptable losses in accuracy, allowing us to interpolate the cooling function in higher dimensions than previously possible.}
However, the loss of distance information associated with the projection step can impact the point selection in later iterations, leading to lower quality simplices.
This might be explored in future work to further improve the algorithm.


\section*{Acknowledgments}
This research was supported by the COMPLEX project from the European Research Council (ERC) under the European Union's Horizon 2020 research and innovation program grant agreement ERC-2019-AdG 860744.

\bibliographystyle{elsarticle-num} 
\bibliography{cas-refs}

\end{document}